\documentclass[sigconf]{acmart}
\renewcommand\footnotetextcopyrightpermission[1]{} 
\usepackage{booktabs} 
\usepackage{float}

\begin{document}
\title{Aiding Autobiographical Memory by Using Wearable Devices}

\author{WANG Jingyi}
\affiliation{%
  \institution{Graduate school of Information, Production and Systems, Waseda University}
  \city{Kitakyushu} 
  \state{Fukuoka} 
  \postcode{808-0135}
}
\email{wangjingyi@toki.waseda.jp}

\author{Tanaka Jiro}
\affiliation{%
  \institution{Graduate school of Information, Production and Systems, Waseda University}
  \city{Kitakyushu} 
  \state{Fukuoka} 
  \postcode{808-0135}
}
\email{jiro@aoni.waseda.jp}








\begin{abstract}
In this paper, we investigate the effectiveness of two distinct techniques (Special Moment Approach \& Spatial Frequency Approach) for reviewing the lifelogs, which were collected by lifeloggers who were willing to use a wearable camera and a bracelet simultaneously for two days. Generally, Special moment approach is a technique for extracting episodic events and Spatial frequency approach is a technique for associating visual with temporal and location information, especially heat map is applied as the spatial data for expressing frequency awareness. Based on that, the participants were asked to fill in two post study questionnaires for evaluating the effectiveness of those two techniques and their combination. The preliminary result showed the positive potential of exploring individual lifelogs using our approaches.
\end{abstract}

%
%



\maketitle

\section{Introduction}

\begin{figure}{}
\includegraphics[height=1.7in, width=3in]{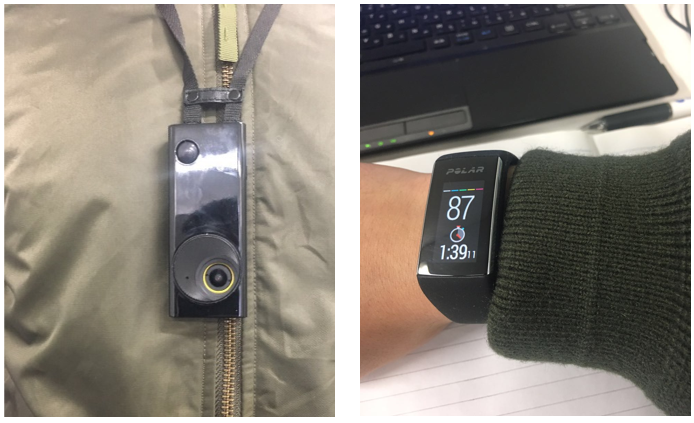}
\caption{The overview Autographer (L) \& Polar A360 (R).}
\label{fig1}
\end{figure}

Autobiographical memory~\cite{conway1990autobiographical} is a memory system consisting of episodes recollected from an individual's life, based on a combination of episodic and semantic memory. As for episodic memory, it mainly contains personal experiences, specific objects, people and events taken place at particular time and space. On the other hand, semantic memory includes general knowledge and facts about the world. It is acknowledged that everyone can hardly remember all events taken place before. In such case, lifelogging, a visualization way of retrieval to memory, is providing us with solutions to supporting human autobiographical memory.\\
\newline
In recent year, due to the development of storage technology, the price of high-capacity storage devices getting cheaper, so people are increasingly easy to get lifelog device to record the daily life, but also can help people to review the past and enhance people's memory, especially for the people who are suffering from memory impairment. However, Autographer~\cite{Autographer} is one of the lifelog devices passively captured images in current research. Due to its nature of collecting various sensory data attached to images, it is applied to capture personal life. Another device involved in this paper is Polar A360~\cite{Polar}, with the photoelectric transmission measurement method, recording the heart rate once per second. Figure.\ref{fig1} illustrates the lifelog devices utilized in this paper.\\
\newline
In the next sections, we will introduce our research in detail. Firstly, general work in lifelog and closely related work will be reported. Then, an algorithm for image preprocessing is applied in our research. After that, the procedure of two approaches will be introduced. The experiment and evaluation for estimating the effectiveness of our approaches will be brought out. Finally, do some brief introduction of our contribution in this field and the conclusion according to the result that we acquired.

\section{Related Work}
In most of the studies using lifelog devices, lifelogs are often displayed in the form of events or other kind of visualization results~\cite{doherty2013wearable,gurrin2014lifelogging,doherty2012experiences,doherty2011automatically}. These events are meaningful logical units that are derived from the merged various sensor data. Doherty et al.~\cite{doherty2012experiences} gave the solutions as events segmentation, events importance and events association to memory aid. Moreover, a web viewer, constructed with the purpose of exploiting the characteristics of personal memory, was reported for supporting the aims of memory rehabilitation~\cite{doherty2011automatically}. \\
\newline
Several researches have been turned out that emotion is crucial to memory construction and retrieval~\cite{reisberg2003memory,singer2010remembered}. However, changes in emotion are usually accompanied by changes in heart rate, which has been explained in the neuroscience field~\cite{appelhans2006heart,nakahara2009emotion}. In particular, Palomba et al.~\cite{palomba1997visual} mentioned the correlation of heart rate and memory. Therefore, Niforatos et al.~\cite{niforatos2015pulsecam} proposed a biophysical driven capture process to acquire events used for aiding user's recoverable memory, but it only reported the idea, prototype and planned study design without any detailed evaluation in that paper. \\
\newline
Prior work reported by Chowdhury et al.~\cite{chowdhury2015my} gave the hints of generating key frames for daily life review. A recent study of his team~\cite{chowdhury2016user} investigated the effectiveness of four distinct techniques used for memory reminiscence. Four distinct techniques represent for (a) visual-temporal; (b) visual-spatial; (c) visual-temporal-spatial; (d) trending location. In his research, it proved that the combination of visual, temporal and spatial information had the best effect on visual lifelogs reviewing. Nevertheless, we introduced heat map as the spatial information, which is more intuitive for users` location awareness and frequency.\\

\section{Research Goals and Approaches}
In general, we try to investigate the powerful and meaningful methods for assisting autobiographical memory. With the insight of prior work~\cite{niforatos2015pulsecam,chowdhury2016user}, we propose two approaches utilizing the biophysical data (heart rate) and lifelogs (images \& GPS) for generating the effective visual retrievable memory. Two approaches, called Special Moment approach and Spatial Frequency approach, are applied in aiding user's daily life reviewing. Besides, we aim at evaluating the effectiveness from the combination of special events and space-time, to prove the feasibility of those approaches.

\section{Robust Motion Deblurring}
However, due to the user's movement or camera swing, it is likely to cause the photo to be blurred. In order to be able to get more and more effective photos, we used the algorithm proposed by Li Xu and Jiaya Jia, called robust motion deblurring~\cite{xu2010two}. After completing the collection of pictures, all the pictures should be preprocessed with this method, get the next step in the need for images. The following fig.\ref{fig2} explains how their algorithm works.

\begin{figure}
\includegraphics[height=1.7in, width=3in]{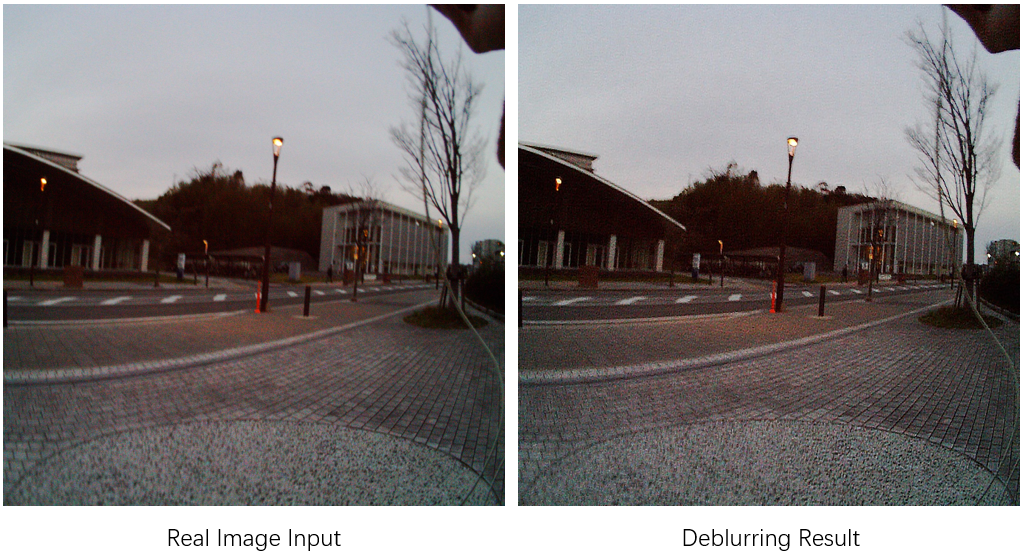}
\caption{Lifelog image input (L) and processed image after deblurring (R).}
\label{fig2}
\end{figure}
\section{Special Moment Approach}

\subsection{Obtain the Heart Rate}

\begin{figure*}
\includegraphics[height=1.8in, width=6.5in]{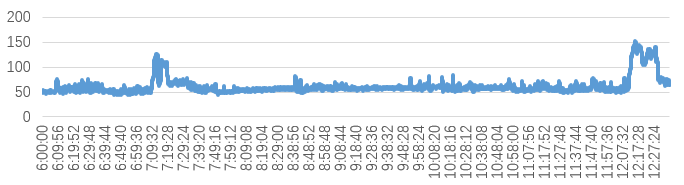}
\caption{Raw heart rate data captured by Polar A360.}
\label{fig3}
\end{figure*}

With the photoelectric transmission measurement method, the sensor in contact with the skin will issue a beam of light hit the skin, measuring the reflected / transmitted light, resulting in heart rate value. Once the Polar A360 is connected to the computer and through the device comes with the synchronization tool Polar Flow, which will synchronize the data to the online database for the user to export or view. Exporting the heart rate into a CSV file is helpful for the later data processing. Figure.\ref{fig3} shows the raw heart rate generated by Excel. In this picture, we only extract part of the heart rate data, from 6:00 am to 12:26 pm, a total of 6 hours and 26 minutes.

\subsection{Normalize the Heart Rate}

\begin{figure*}
\includegraphics[height=1.8in, width=6.5in]{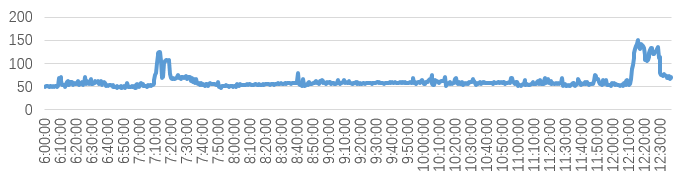}
\caption{Heart rate data after normalization.}
\label{fig4}
\end{figure*}

As it is known, Polar A360 records heart rate every second, but taking into account that the Autographer takes a photo every 30 seconds. So in order to link heart rate and pictures together, calculate the average of every 30 seconds of heart rate data as one value, corresponding to each picture. Figure.\ref{fig4} illustrates the result of heart rate normalization. It is apparent that the curve becomes more sparse compared to the foreahead one, which counts a lot in extracting part.

\subsection{Combine the Heart Rate and Images}

\begin{figure*}
\includegraphics[height=3in, width=6.5in]{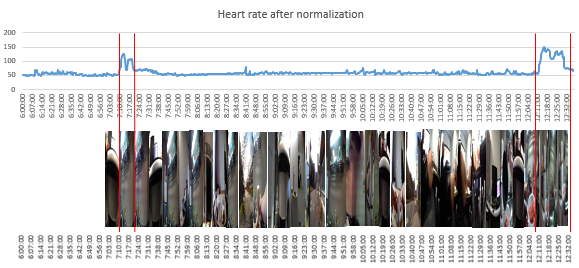}
\caption{Associate the heart rate with the images.}
\label{fig5}
\end{figure*}

According to the second step, we have converted the heart rate data into a 30-second data, for better match with each of the corresponding pictures. Since each image and each set of heart rate data have the time information, we synchronize them with time information as the base point. Figure.\ref{fig5} depicts how the heart rate and image are combined and screened out of the extracting of the part. In this figure, the highlight parts explain the prompt changes of heart rate, which also indicates the processing part.

\subsection{Extract Episodic Events}

\begin{figure*}
\includegraphics[height=4in, width=6.5in]{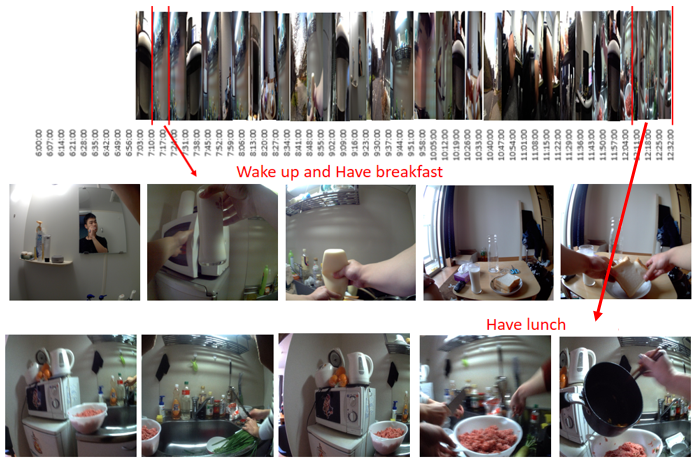}
\caption{Extract episodic events.}
\label{fig6}
\end{figure*}

After finishing the fusion of heart rate and images, it is obvious to extract the periods that user's heart rate changes suddenly. The following pictures are the partial of extracting periods and have been described by the user. In the figure.\ref{fig6}, it is highlighted that only two parts are processed, the definition of the events given by user himself. However, those periods have been extracted manually so far.

\section{Spatial Frequency Approach}

\subsection{Get GPS Information}

\begin{figure*}
\includegraphics[height=2.3in, width=6.5in]{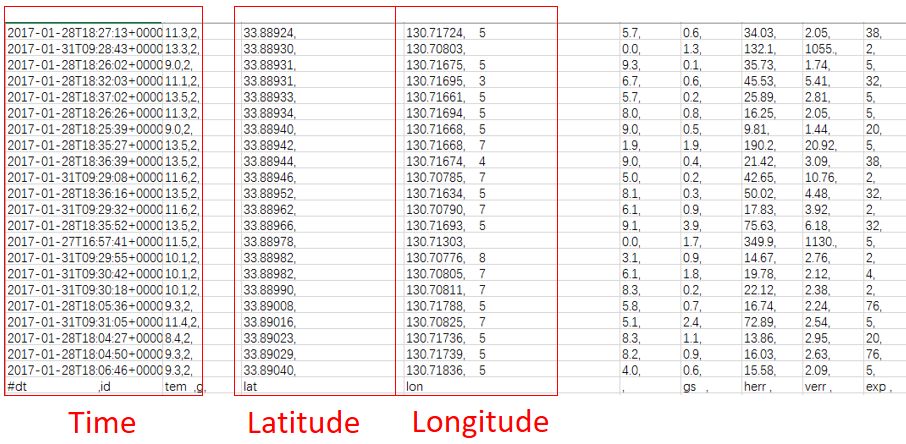}
\caption{The screenshot of excel file of GPS information provided by Autographer.}
\label{fig7}
\end{figure*}

Basically, the GPS information are provided by Autographer. Due to its mechanism, the spatial information is likely to be not acquired in some conditions, especially in the buildings. Thus, the data is not covered all the trails of one's whole recorded day. Despite of that, we intend to apply all available GPS information into the next procedure. In the figure.\ref{fig7}, it contains several information recorded by various embedded sensors. From the left to right, the first one represents for temporal data, and the rest are latitude and longitude.

\subsection{Generate Heat Map}

\begin{figure}
\includegraphics[height=1.7in, width=3in]{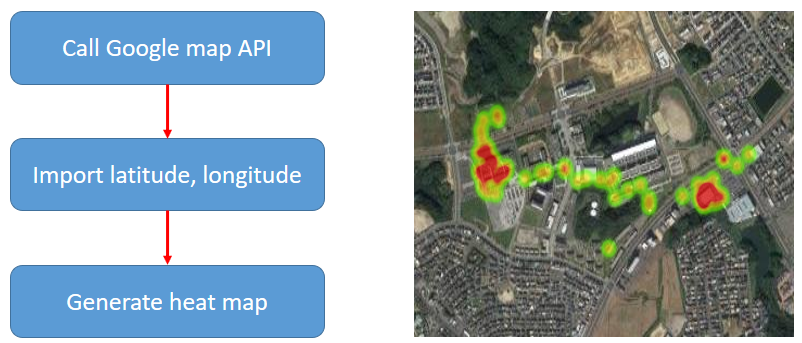}
\caption{Procedure of generating heat map (L) and its result (R).}
\label{fig8}
\end{figure}

Once the GPS data has been obtained, it should be imported to generate heat map by calling the Google Map API~\cite{Heatmap1}. The principle of its generation is simply summarized as four steps:\\
\newline
(a) Set a radius for discrete points to create a buffer; \\
(b) For each discrete point of the buffer, the use of progressive grayscale (the complete gray scale is 0 ~ 255) from the inside out, from shallow to deep fill; \\
(c) As the gray value can be superimposed (the larger the value is, the lighter the color is in the grayscale.) In practice, you can select any channel in the ARGB model as a superimposed gray value, so that for a buffer Area, you can stack the gray value. In such case, the more the buffer cross, the greater the gray value is, meanwhile, this area will be more "hot"; \\
(d) The color is mapped from a 256-color ribbon (such as a rainbow) with the grayscale value after the overlay, and the image is re-colored to achieve the heat map~\cite{Heatmap2}.\\
\newline
It can be simple to aggregate large amounts of data through the heat map, which use a progressive ribbon ensures the elegant performance. Ultimately, it can be very intuitive to show the density of spatial data or frequency level. Figure.\ref{fig8} explains the flow chart of generating heat map and shows how heat map looks like respectively.

\subsection{Match each spot with image}

As aforementioned, heat map is comprised of massive discrete spots. Additionally, each spot is generated with its temporal information, and each image is recorded with its temporal information as well. In such case, it is apparent that each spot corresponds to one image respectively if that image gets GPS information at first. As a matter of fact, only the temporal information of spot and image is the same, they are associated with each other literally.

\subsection{Review the Lifelog}

\begin{figure}
\includegraphics[height=1.7in, width=3in]{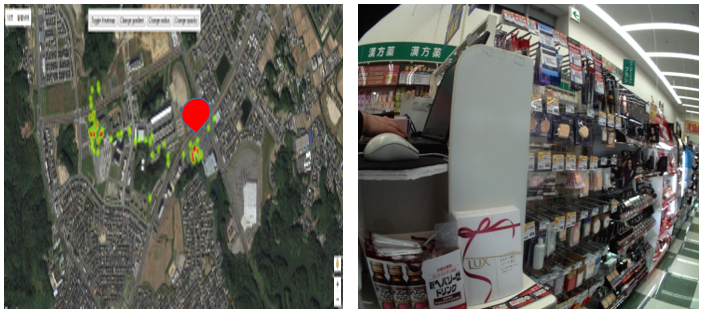}
\caption{Review the lifelog in heat map.}
\label{fig9}
\end{figure}

After all the spots are linked with the corresponding images together, the matching results will be displayed in the web viewer with the combination of GPS spots and images, providing users with visual review to assist autobiographical memory. Figure.\ref{fig9} exemplifies the spot which the user is reviewing and the right picture is according with that. However, it is just the prototype of reviewing lifelogs, remaining several aspects to be improved.

\section{Experiment and Evaluation}

\subsection{Experiment}

\begin{table}[]
\centering
\caption{Subject Demographic Information}
\label{tab:subj}
\begin{tabular}{|l|c|}
\hline
Participants           & 3 males, 1 female \\ \hline
Age                    & 22-24; Mean:23    \\ \hline
Professional           & Students          \\ \hline
Time spent in the city & More than 1 year  \\ \hline
\end{tabular}
\end{table}

The objective of our experiment is to investigate the effectiveness of special moment approach and spatial frequency approach for supporting autobiographical memory. Of course we have also compared to the regular review of utilizing the viewer comes with Autographer.\\
\newline
As Table~\ref{tab:subj} shown, there are 4 participants involved in our research so far, but the number of participants will be increased in the future. And all the participants are students with computer sicience background, living in this city for more than one year.\\
\newline
In the experiment, each participant was asked for using the devices simultaneously for 2 days around 8 hours per day. And then submit the data for processing. After that, all the participants were involved in the personal interview with lifelogs which are processed. Meanwhile, 4 people were required to fill in two questionnaires when looking through their lifelogs. Once the participants were browsing their processed recorded, they were asked to answer two questionnaires (Q1 \& Q2) capturing information regarding: 
\newline
(Q1.1) How meaningful is the presented information? \\
(Q1.2) Is the information presented clearly and can be easily interpreted? \\
(Q1.3) Does the visulization help the lifeloggers in calling up the past in a intuitive way?\\
\newline
(Q2.1) How many events that you can reminisce without any help? \\
(Q2.2) After using Special Moment approach, how many new events that come to your mind? \\
(Q2.3) After using Spatial Frequency approach, how many new events that come to your mind?\\
\newline
Eventually, the evaluation is given out by analysising the feedback of contributors.\\

\subsection{Evaluation}

After collecting the results given by the participants, the evaluation of two approaches for aiding autobiographical memory can be carried out. The evaluators were asked to rate on a Likert Scale ranging from 1 to 5. Therefore, results of the assessment are based on the user's questionnaire, calculating the average of each answers and illustrating the following fig.\ref{fig10}.\\
\newline
It can be seen from the fig.\ref{fig10} that significant changes have taken place in all questions, compared to the Regular Review. Here, Regular Review stands for the participants review their lifelogs by using Autographer application. However, owing to Special Moment Review and Spatial Frequency Review have different impacts on the user's memory recalling in different circumstances, so in the histogram to the three questions corresponding, there are some slight difference between those two methods. In general, the estimation indicates that the approaches enhance the effectiveness of assisting autobiographical memory but respectively.\\
\begin{figure*}
\includegraphics[height=2.3in, width=6.5in]{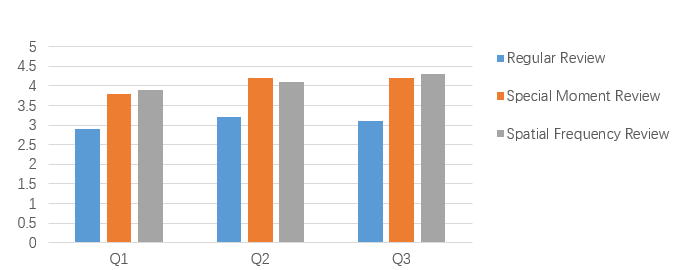}
\caption{Associate the heart rate with the images.}
\label{fig10}
\end{figure*}

\begin{figure*}
\includegraphics[height=2.3in, width=6.5in]{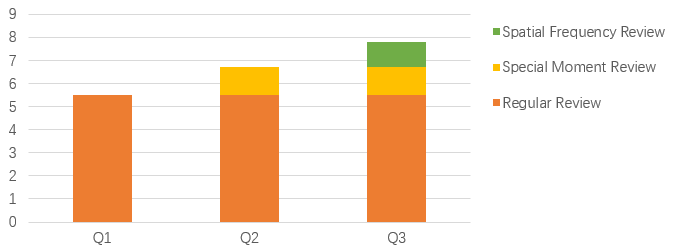}
\caption{Associate the heart rate with the images.}
\label{fig11}
\end{figure*}
In fig.\ref{fig11}, Regular Review explains the basic level of exploring lifelogs as well. As it shown in the second bar, the increase number of new events after using Special Moment approach. The same situation to the last one, which not only reflects the growth of new events after exploiting Spatial Frequency approach, but also the total increase of new events after utilizing two methods literally.

\section{Research Contribution}
In this paper, we firstly brought out the combination of biophysical data and visual-temporal-spatial for reminiscing visual lifelogs. The approaches are unlike to the prior works, for example, we extracted the events in the light of heart rate suddenly changed instead of biophysically driven life logging [12]. Moreover, heat map has been introduced as the spatial information, which is more intuitive for location density and frequency compared to the Chowdhury's work [14]. It has been proved that the potential of exploiting the body signals in memory retrieval, and there will be a possibility to introduce other closely related body signals for personal lifelogs reminiscence.

\section{Conclusion}
In summary, this paper explains the combination of two distinct visualization techniques supporting explore personal lifelogs and recall the past, which make the reviewing engaging and informative using Autographer and Polar A360 simultaneously per day. All captured images are engaged with heart rate and Spatial-Temporal information, making the visualization of memory recalling more informatively. Moreover, the results with respect to the effectiveness of assisting autobiographical memory has been discussed, suggesting the meaningful and easily interpreted summarization of the lifelogs using Special Moment Approach and Spatial Frequency Approach. The techniques reported in this paper explicitly set forth the visualization fusion of biophysical data and GPS brought out the feasibility of making more available memory retrieval. The results also indicate the positive potential compared to the prior work. We are completely convinced that it will show the much better results and significant changes when the quantities of participants increase.

\bibliographystyle{ACM-Reference-Format}
\bibliography{sigproc} 

\end{document}